\documentclass[runningheads]{llncs}

\usepackage{amsmath, amssymb, amsfonts}
\usepackage{algorithm}
\usepackage{algpseudocode}
\usepackage{graphicx}
\usepackage{hyperref}
\usepackage{enumitem}
\usepackage{tikz-cd}
\usetikzlibrary{arrows.meta,positioning,calc,fit,backgrounds}
\usepackage{multirow}
\usepackage{adjustbox}

\usepackage{bm}
\usepackage{array}
\usepackage{tabularx}
\usepackage{booktabs}
\usepackage{placeins}
\usepackage[T1]{fontenc}
\usepackage{float}

\hypersetup{
    colorlinks=true,
    linkcolor=black,
    urlcolor=blue
}

\begin{document}

\title{Active Inference for Interaction-Mediated Control of a High-Dimensional Robotic Arm}
\titlerunning{Active Inference for Interaction-Mediated Control}

\author{Fraser C. Paterson\inst{1}\orcidID{0009-0003-4392-7092} \and
Sebastian Stein\inst{1}\orcidID{0000-0003-1828-4008} \and
Markus Klar\inst{1}\orcidID{0000-0003-2445-152X} \and
John H. Williamson\inst{1}\orcidID{0000-0001-8085-7853} \and
Roderick Murray-Smith\inst{1}\orcidID{0000-0003-4228-7962} }

\authorrunning{F. C. Paterson et al.}

\institute{University of Glasgow, Glasgow, Scotland, UK\\
\email{3103294P@student.gla.ac.uk}}

\maketitle

\begin{abstract}
We propose interaction-mediated control via Active Inference as a general architectural approach to high-dimensional control problems in Human--Computer Interaction (HCI). This architecture recasts user interaction as the provision of evidence about a latent task objective, rather than the direct specification of plant-control inputs. The mediating function is distributed between an interaction broker, which selects informative user queries and performs Bayesian inference over user preferences, and an Active Inference controller, which autonomously plans and acts under the resulting preference information to control the plant. This division of labour decouples the semantics of user interaction from those of low-level plant control. We instantiate the architecture in a simulated, multi-link robotic arm to perform a simultaneous whole-arm target-coverage task. A simulated user communicates exclusively through a clutch-style binary evaluative channel, without specifying joint-torque commands. Across increasing arm dimensionalities, the architecture achieves successful interaction-mediated control, although task success is lower than when the controller receives the true target preferences directly. Exact target-subset identification also remains imperfect, highlighting the distinction between preference inference and successful task completion. The experimental findings provide an initial computational demonstration of the proposed architecture under controlled, matched-model assumptions and motivate its further investigation in broader HCI applications.

\end{abstract}

\section{Introduction}
\label{sec:introduction}

Users of interactive, computational systems often express their intentions and preferences at a higher semantic level than the space of controls required to realise them. ``Tie my shoelaces'' is a concise, abstract imperative, over and above all the constituent fine motor control commands that will constitute its execution. In consequence, there is a fundamental distinction between the specification of \textit{what} a system should achieve and the determination of \textit{how} it should be achieved. Requiring users to directly interface with the low level controls can become burdensome when those controls involve many dynamically coupled variables or if they cannot be exposed through a tractable interface compatible with the limited bandwidth of human interaction \cite{ZHENG2025192}. This problem is especially prominent in robotic, assistive, and shared-autonomy settings, where low-bandwidth user input must be reconciled with comparatively high-dimensional control \cite{muelling2017autonomy,javdani2015shared,gopinath2020active}. We address this problem through \emph{interaction-mediated control via Active Inference}, in which user input communicates evidence about their latent, task-relevant preferences, as opposed to direct control. A mediating architecture infers the user's preferences from interaction and communicates the resulting user-preference information to an autonomous, active inference controller in direct control of the plant. We propose an explicit division of labour within the mediating system, separating an Interaction Broker Agent (IBA), responsible for querying and inferring user preferences, from a Low--Level Control Agent (LLCA), responsible for plant control, under the resulting preference information. Section~\ref{sec:interaction-mediated-control-aif} develops this architectural proposal.

Our principal contribution is the formal specification and computational realisation of this interaction-mediated architecture. We instantiate our approach in a simulated multi-link robotic arm and evaluate whether a clutch-style, binary evaluative user-interaction channel can convey useful information about a simulated user's latent target objective, without requiring the user to specify joint-torque controls. The resulting experiments provide an initial demonstration of the architecture and motivate its investigation in broader HCI applications.


\section{Interaction-Mediated Control via Active Inference}
\label{sec:interaction-mediated-control-aif}

\subsection{Mediated Interaction as a General Architectural Response}
As opposed to a direct mapping from user input into low--level plant controls, a \emph{mediating mechanism} may stand between the user and the plant. We refer to this general arrangement as \emph{interaction-mediated control}. The mere fact of mediation does not prescribe a particular semantics for user input. For example, user input might retain the semantics of a \textit{control} command, expressed through a lower-dimensional latent control space afforded by the mediating mechanism. An alternative semantics is to treat user input as evaluative evidence about whether the mediator's proposed behaviour accords with the user's task objective. This alternative decouples the semantics of user interaction from the semantics of Low--Level plant actuation. We adopt this latter semantics: user interaction provides evidence about the user's latent task-preference, while the mediating system determines the plant actions through which the inferred objective may be realised.

\subsection{Active Inference as an Interaction-Mediating Mechanism}
We propose Active Inference (AIF) as a framework for instantiating this evaluative form of interaction-mediated control \cite{friston2017active}. Rather than treating user input as a direct control command, an Active-Inference mediator maintains probabilistic beliefs about latent user preferences. User-facing queries serve the epistemic role of soliciting user-feedback expected to reduce uncertainty about user-preferences. The resulting preference information can subsequently guide pragmatic, plant-facing actions, directed towards the realisation of the inferred objective. This provides a common, inferential perspective on the related issues of learning \textit{what} the user wants in addition to \textit{how} these preferences can be realised.

Recent work on Active Inference for HCI has similarly identified an online ``transduction'' configuration in which an Active Inference interface lies between a user and an existing system and mediates information flow between them \cite{murraysmith2025active}. We refer to the resulting arrangement as \emph{interaction-mediated control via Active Inference}. Active Inference has also been investigated for robotic body perception and reaching \cite{oliver2022empirical}, and for robot-arm torque control \cite{meo2023adaptation}. Our focus here is on coupling autonomous plant control to preference information inferred through user interaction. Previous work has also employed Active Inference to model mouse pointing and clicking in continuous state, observation, and action spaces \cite{klar2026pointing}. The present work instead investigates a mediated architecture in which a simulated user's latent objective is inferred from evaluative interaction before autonomous plant control.

\subsection{Decomposing the AIF-Mediator: Inferential Division of Labour}
The proposed mediating function exists to solve two distinct problems: inference over \emph{what} the user wants and determining \emph{how} the inferred objective can be realised through autonomous plant control. We make this division explicit by decomposing the mediating function into an Interaction Broker Agent (IBA) and Low--Level Control Agent (LLCA). The IBA queries the user and performs exact Bayesian inference over latent user preferences. The LLCA performs autonomous plant control under the resulting posterior-marginal preference information supplied by the IBA. We simulate the user side of the interaction and refer to this entity as the Human/User Agent (HUA) which has private access to the task objective and provides evaluative feedback to the IBA. The resulting architecture is deliberately heterogeneous: its components perform distinct, local computations and communicate through restricted interfaces rather than forming a single, monolithic inference-and-control mechanism.

\begin{figure}[h]
    \centering
    \includegraphics[width=0.9\linewidth]{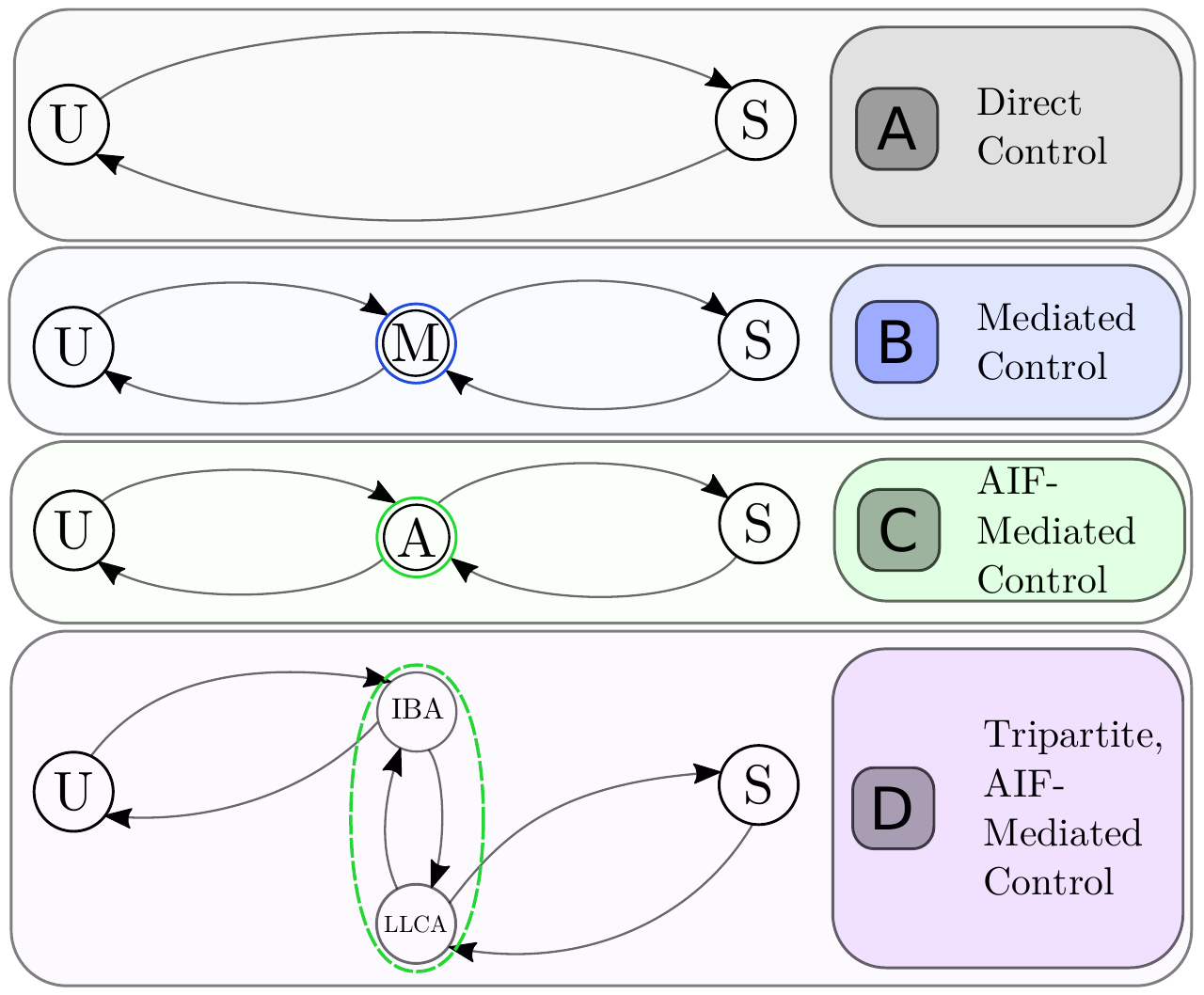}
    \caption{Four arrangements illustrate the progression from \textbf{(A)} direct user--system control to \textbf{(B)} interaction-mediated control, \textbf{(C)} Active Inference-mediated control, and \textbf{(D)} its distributed realisation through the HUA, IBA, and LLCA. In (D), the IBA and LLCA jointly constitute the mediating mechanism.}
    \label{fig:example-fig-1}
\end{figure}


\section{AIF Interaction-Mediated Control for a Multi--Link Arm}
\label{sec:architecture}

\subsection{Task Definition}

We consider a simultaneous whole-arm coverage task for a planar robotic arm comprising $N$ equal-length links. The configuration of the arm is given as the vector of angles: $\boldsymbol{\theta}=(\theta_1,\ldots,\theta_N)\in\mathbb{T}^N$, where $\theta_1$ is measured relative to the fixed base frame and subsequent angles relative to the preceding link. Low--Level actions are per-joint torques $\boldsymbol{\tau}\in\mathbb{R}^N$. The arm is situated inside a workspace which contains $K$ circular objects $\mathcal{O}=\{O_1,\ldots,O_K\}$. The private target subset is represented by $z^\ast\in\{0,1\}^K$, with $z_k^\ast=1$ indicating that object $O_k$ is a target. Let $\mathcal{L}_i(\boldsymbol{\theta})\subset\mathbb{R}^2$ denote the region occupied by arm-link $i$ and $\mathcal{A}(\boldsymbol{\theta})=\bigcup_i\mathcal{L}_i(\boldsymbol{\theta})$ be the geometry of the whole arm. Object $O_k$ is covered when $\mathcal{A}(\boldsymbol{\theta})\cap O_k\neq\emptyset$, and the task succeeds when all target objects are covered \textit{simultaneously}. Non-target coverage is not penalised and no unique arm pose is prescribed; the LLCA determines how the target objective is realised through joint torques.

\subsection{Local Models, Objectives, and Interface Contracts}
The three interacting components are functionally heterogeneous and implement distinct inference, evaluation, and control processes. We therefore specify the minimal local model, operative computation, and interface contract of each component. Full implementation details are given in Appendices~\ref{app:hua}--\ref{app:llca}. Fig.~\ref{fig:example-fig-2} depicts one whole interaction cycle. 

\subsubsection{Human/User Agent}
The simulated HUA privately knows $z^\ast$. At outer cycle $t$, it receives the LLCA-generated rollout-pool context $B_{t,\mathrm{all}}$ and IBA-selected query index $m_t^{\mathrm{query}}$, and samples
\begin{equation}
    u_t^{\mathrm{preview}}
    \sim
    P_B
    \left(
        \cdot
        \mid
        z^\ast,
        B_{t,\mathrm{all}},
        m_t^{\mathrm{query}}
    \right),
    \qquad
    u_t^{\mathrm{preview}}\in\{0,1\}.
\end{equation}
Relative to the current rollout-pool context, $u_t^{\mathrm{preview}}=0$ denotes satisfaction with the selected preview and $u_t^{\mathrm{preview}}=1$ dissatisfaction. The HUA communicates neither $z^\ast$ nor motor commands and is not implemented as an Active Inference agent. Consequently, the HUA performs no VFE minimisation or EFE planning. The response model $P_B$ is specified in Appendix~\ref{app:hua}.

\subsubsection{Interaction Broker Agent}
The IBA treats the HUA's private target indicator as a latent variable $z\in\mathcal{Z}$, where $\mathcal{Z}$ is the finite set of admissible target-subset hypotheses. At outer interaction cycle $t$, it maintains a categorical pre-query belief $q_t^{\mathrm{pre}}(z)$ over this hypothesis space. Given the LLCA-generated rollout-pool context $B_{t,\mathrm{all}}$, the IBA evaluates candidate preview queries using the same pool-contextual HUA response-model family $P_B(u\mid z,B_{t,\mathrm{all}},m)$ introduced above. The IBA selects the candidate query expected to provide the greatest information about the latent target subset, instantiating an epistemic objective related to Bayesian optimal experimental design \cite{sajid2022active}:
\begin{align}
    \mathrm{IG}_t^{\mathrm{IBA}}(m)
    &=
    I
    \left(
        z;
        u_t^{\mathrm{preview}}
        \mid
        B_{t,\mathrm{all}},
        m,
        q_t^{\mathrm{pre}}
    \right),
    \\
    m_t^{\mathrm{query}}
    &=
    \arg\max_m
    \mathrm{IG}_t^{\mathrm{IBA}}(m).
\end{align}

After receiving the HUA's binary evaluation $u_t^{\mathrm{preview}}$, the IBA updates its belief via direct application of Bayes' rule:
\begin{equation}
    q_t^{\mathrm{post}}(z)
    \propto
    q_t^{\mathrm{pre}}(z)
    P_B
    \left(
        u_t^{\mathrm{preview}}
        \mid
        z,
        B_{t,\mathrm{all}},
        m_t^{\mathrm{query}}
    \right),
    \qquad
    z\in\mathcal{Z}.
\end{equation}
The resulting posterior is normalised over $\mathcal{Z}$. The IBA does not perform VFE minimisation for this update. Rather than communicating either the full categorical belief or its MAP target hypothesis to the LLCA, the IBA marginalises over target hypotheses to obtain object-wise preference vectors:
\begin{equation}
    \mathbf{p}_t^{\mathrm{IBA},s}
    =
    \begin{bmatrix}
        p_{t,1}^{\mathrm{IBA},s} &
        \cdots &
        p_{t,K}^{\mathrm{IBA},s}
    \end{bmatrix}^{\top}
    =
    \mathbb{E}_{q_t^s}[z],
    \qquad
    s\in\{\mathrm{pre},\mathrm{post}\}.
    \label{eq:iba_target_beliefs}
\end{equation}
Each component $p_{t,k}^{\mathrm{IBA},s} = P_{q_t^s}(z_k=1)$ is the marginal probability that workspace object $O_k$ belongs to the HUA's target subset. The pre-feedback vector $\mathbf{p}_t^{\mathrm{IBA},\mathrm{pre}}$ is supplied to the LLCA for query-rollout generation, whereas the post-feedback vector $\mathbf{p}_t^{\mathrm{IBA},\mathrm{post}}$ is supplied for subsequent execution planning. The IBA itself issues no torque commands and does not select an execution rollout. Full likelihood, expected-information-gain, and posterior-update expressions are given in Appendix~\ref{app:iba}.

\subsubsection{Low--Level Control Agent}
The LLCA is the exclusive controller\footnote{The LLCA's generative model is implemented in JAX \cite{jax2018github}.} of the multi-link arm plant. The LLCA observes all joint angles through the coordinate vector $o_t\in\mathbb{R}^N$, corresponding to the unmodified MuJoCo joint positions representing the physical configuration $\theta_t\in\mathbb{T}^N$. No stochastic observation noise is added to this observation stream, although the LLCA represents observation uncertainty internally through its likelihood model. See Appendix~\ref{app:llca} for details. Joint angular velocities, Cartesian geometry, and task-state variables are omitted from the LLCA observation modality. The LLCA receives an object-wise preference vector \(\mathbf{p}_t\in[0,1]^K\) from the IBA, where \(\mathbf{p}_t=\mathbf{p}_t^{\mathrm{IBA},\mathrm{pre}}\) during query planning and \(\mathbf{p}_t=\mathbf{p}_t^{\mathrm{IBA},\mathrm{post}}\) during execution planning. It issues joint-torque control actions \(\tau_t\in\mathbb{R}^N\) to the plant. Internally, the LLCA represents\footnote{If variable $x$ belongs to the plant/generative process, we use a ``hat'' to denote the corresponding variable in the agent's generative model: $\hat{x}$.} the latent arm state using Euclidean joint-coordinate representations of the joint angles and their angular velocities, maintaining Gaussian beliefs over this state representation: $\hat{s}_t^\top =
\begin{bmatrix}
    \hat{\theta}_t^\top &
    \hat{\dot{\theta}}_t^\top
\end{bmatrix}
\in \mathbb{R}^{1 \times 2N}$. 
The LLCA maintains a generative model of the arm's dynamics, which it uses to predict the evolution of the latent arm state. At each filtering step, the LLCA first propagates its current Gaussian state belief forward under this internal dynamical model to obtain a predictive prior $q_t^{\mathrm{LLCA}}(\hat{s}_t)$. It subsequently updates its predictive prior belief into the variational posterior $q_t^{\mathrm{LLCA}}(\hat{s}_t \mid o_t)$ by performing variational inference on the observed joint-angle configuration $o_t$. Specifically, the variational posterior is obtained by minimising the variational free energy
\begin{align}
    \mathcal{F}_t^{\mathrm{LLCA}}
    &=
    \mathbb{E}_{q_t^{\mathrm{LLCA}}(\hat{s}_t \mid o_t)}
    \left[
        -\log p(o_t\mid\hat{s}_t)
    \right]
    \nonumber\\
    &\qquad
    +
    D_{\mathrm{KL}}
    \left(
        q_t^{\mathrm{LLCA}}(\hat{s}_t \mid o_t)
        \,\|\,
        q_t^{\mathrm{LLCA}}(\hat{s}_t)
    \right). \label{eq:vfe}
\end{align}

Predictive-prior propagation is performed using an Unscented Transform, whereas the subsequent perceptual correction is obtained by minimising this variational free energy with respect to the parameters of the Gaussian posterior. Motor actions are selected using receding-horizon random-shooting control. For each planning call, the LLCA samples $M$ candidate finite-horizon torque sequences. We denote the resulting candidate-plan pool by
$\Pi_t
=
\{
    \pi_t^{(0)},
    \ldots,
    \pi_t^{(M-1)}
\}$, where each candidate plan 
$\pi_t^{(m)}
=
(\hat{\tau}_{t,1}^{(m)},
\ldots,
\hat{\tau}_{t,H}^{(m)})$
contains $H$ planned torque vectors. Each candidate torque sequence is propagated from the LLCA's current Gaussian state belief through its internal dynamical model, yielding the corresponding predicted state-belief trajectory. We use the term ``rollout pool'' to refer to these $M$ candidate plans together with their predicted consequences. When the role of a particular pool must be distinguished, we write $\Pi_t^{\mathrm{query}}$ for a pool used for query construction and $\Pi_t^{\mathrm{exec}}$ for a pool used for action selection.

The object-wise preference vector $\mathbf{p}_t$ parameterises an embodied preference over whole-arm coverage. Let $\phi(\vartheta)\in[0,1]^K$ denote smooth, object-wise, whole-arm coverage features for a joint-angle coordinate vector $\vartheta\in\mathbb{R}^N$. We define the LLCA preference density as
\begin{equation}
    \log
    P_C
    \left(
        \phi(\vartheta)
        \mid
        \mathbf{p}_t
    \right)
    =
    -
    \frac{1}{2\sigma_{\phi}^{2}}
    \sum_{k=1}^{K}
    p_{t,k}
    \left(
        1-\phi_k(\vartheta)
    \right)^2
    +
    \mathrm{const}. \label{eq:llca_preference}
\end{equation}
Thus, each IBA marginal $p_{t,k}$ directly weights the LLCA's preference for covering object $O_k$; no MAP target subset is extracted before planning. For candidate rollout $\pi$, the pragmatic value at planning step $h$ is
\begin{equation}
    \mathrm{PV}_h(\pi)
    =
    \mathbb{E}_{
        q^{\mathrm{LLCA}}(\hat{s}_h\mid\pi)
    }
    \left[
        \log
        P_C
        \left(
            \phi(\hat{\theta}_h)
            \mid
            \mathbf{p}_t
        \right)
    \right]. \label{eq:pragmatic_value}
\end{equation}

In the reported implementation, this expectation is approximated by evaluating the log preference at the joint-angle component $\mu_{\theta,h}^{\pi}$ of the predicted Gaussian mean. The exact implementation is specified in Appendix~\ref{app:llca}. In all experiments reported in this paper the LLCA's computation of information gain is disabled. Consequently, rollout selection is driven solely by whole-horizon pragmatic value. For the post-feedback execution planning call
\begin{equation}
    \pi_t^\star
    =
    \arg\max_{\pi \in \Pi_t^{\mathrm{exec}}}
    \frac{1}{H}
    \sum_{h=1}^{H}
    \mathrm{PV}_h(\pi).
\end{equation}
The first planned torque vector in the selected execution rollout is clipped to the plant actuator bounds. The resulting torque is held fixed and repeatedly applied for the configured action-repeat block before the next round of planning and action selection. Subsequent torque vectors in the planned rollout are discarded before replanning. During the pre-query planning call, the LLCA additionally exposes the unweighted object-wise terminal basis pool $B_{t,\mathrm{all}}$ to the IBA for query selection. The LLCA does not itself select the HUA query, infer $z^\ast$, or receive an execution command from the IBA. Full dynamical-model, state-inference, rollout-generation, coverage, and planning details are given in Appendix~\ref{app:llca}.

\subsection{Complete Outer Interaction Cycle}


The plant does not advance during query construction, HUA evaluation, or IBA belief updating; physical action follows the second LLCA planning call (Fig.~\ref{fig:example-fig-2}). Algorithm~\ref{alg:outer-cycle} in Appendix~\ref{app:interalg} provides pseudocode for the complete interaction cycle.
\begin{enumerate}
    \item \textbf{Query planning.}
    The IBA supplies $\mathbf{p}_t^{\mathrm{IBA},\mathrm{pre}}$ to the LLCA, which generates the query rollout pool $\Pi_t^{\mathrm{query}}$ and its terminal basis $B_{t,\mathrm{all}}$.

    \item \textbf{Query selection and evaluation.}
    The IBA selects $m_t^{\mathrm{query}}$ by expected information gain. The HUA returns the binary evaluation $u_t^{\mathrm{preview}} \in \{0, 1\}$.

    \item \textbf{Posterior inference and preference communication.}
    The IBA updates its categorical belief to $q_t^{\mathrm{post}}(z)$ and marginalises this posterior to obtain $\mathbf{p}_t^{\mathrm{IBA},\mathrm{post}}$.

    \item \textbf{Replanning and execution.}
    Under the updated preference information, the LLCA independently generates a fresh execution pool $\Pi_t^{\mathrm{exec}}$, selects a rollout, and executes only its first torque.
\end{enumerate}

\begin{figure}[h]
    \centering
    \includegraphics[width=1.0\linewidth]{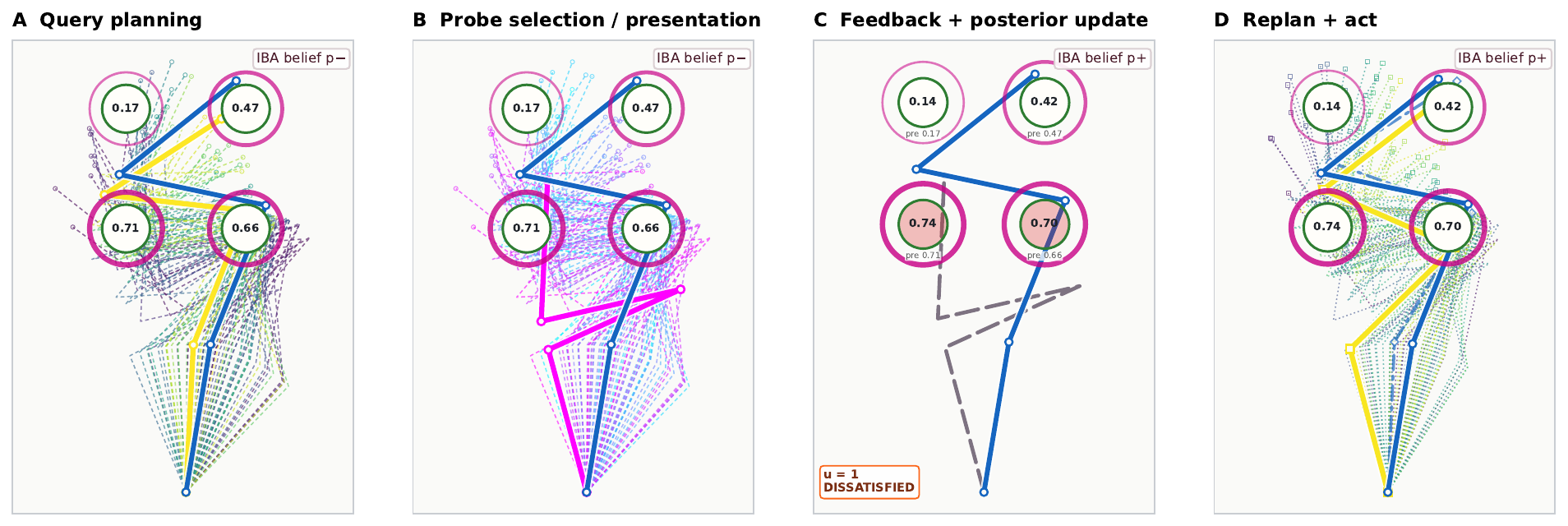}
    \caption{An outer interaction cycle at $t=111$ from an illustrative simulation demonstrates query selection, feedback assimilation, and execution. \textbf{(A)} The LLCA generates a query rollout pool, coloured by pragmatic-value rank. The solid yellow rollout has the highest pragmatic-value. \textbf{(B)} The IBA ranks the same pool by expected information gain (measured in nats) and selects a query, which is not executed. The solid pink rollout has the greatest expected information gain. \textbf{(C)} User feedback (dissatisfaction $u_t^{\mathrm{preview}}=1$, in this case) allows the IBA to update its posterior. Red-shaded interiors identify the HUA's private targets. \textbf{(D)} The LLCA generates a fresh execution pool and applies the selected plan's first torque over the action-repeat block. Object-centred values and magenta rings represent target marginals; A and D share a pragmatic-value ranking scale. The solid blue arm remains fixed in A--C; the dash-dotted blue arm in D shows the realised post-action pose.}
    \label{fig:example-fig-2}
\end{figure}

The queried rollout is an epistemic probe rather than an execution command. The IBA carries its posterior into the next cycle as $q_{t+1}^{\mathrm{pre}}(z)=q_t^{\mathrm{post}}(z)$.


\section{Experimental Design}
\label{sec:experimental-design}

\subsection{Experimental Question and Task}
\label{sec:task}
We evaluate the interaction-mediated control architecture in silico using the MuJoCo-based generative process implemented via \texttt{dm\_control} \cite{tunyasuvunakool2020dmcontrol}. We use the simultaneous whole-arm coverage task defined in Section~\ref{sec:architecture}, with $K=4$ circular workspace objects arranged in a $2\times2$ grid and exactly two target objects per episode. All six possible two-object target subset hypotheses are evaluated. The principal experimental variable is the arm's dimensionality (number of links: $N$, from $3$ to $6$). For each $N$, all six target subsets are evaluated over $24$ repeated episodes, yielding $144$ episodes per arm dimensionality and $576$ episodes per experimental condition.

\subsection{Matched Experimental Conditions}
\label{sec:conditions}
An LLCA-only parameter search selected one configuration per arm dimensionality. For $N=3,4,5,6$, the respective $(M,H,\mathrm{AR})$ settings were $(64,2,25)$, $(64,2,15)$, $(64,4,15)$, and $(32,4,15)$, where $M$ is the number of candidate plans, $H$ the temporal depth per candidate plan, and $\mathrm{AR}$ the action-repeat factor. The isotropic process-noise covariance was fixed at $Q=10^{-3}I_{2N}$ for all arm dimensionalities.. These settings are held fixed across both experimental conditions.

In the direct-preference reference condition, the HUA--IBA interaction loop is disabled and the true object-wise target preference is directly supplied to the LLCA. This condition assesses controller task competence under direct access to the objective. The interaction-mediated condition is the primary experimental condition. In this setting, the HUA alone has direct access to the true target subset $z^\ast$. The IBA infers target preferences from binary HUA feedback and communicates posterior-marginal object preferences to the LLCA, as specified in Section~\ref{sec:architecture}. The LLCA receives no direct target information and independently plans under the communicated marginals. Task and controller settings are otherwise matched. The simulated HUA uses the same pool-contextual response-model family as the IBA: this is a controlled matched-model comparison, not an evaluation of real human behaviour or user-model mismatch. 

\subsection{Evaluation}
\label{sec:evaluation}

We report the task success rate and time-to-first-success.\footnote{The source code and experimental configurations are publicly available at \url{https://github.com/FraserP117/DIFAI-Arm-IWAI2026}.} An episode succeeds when the arm simultaneously covers all target objects at least once before termination. Success is checked after every generative-process step. Time-to-first-success is reported as $T_{\mathrm{gp}}$, the zero-based index of the first successful generative-process step, and $T_{\mathrm{outer}}$, the zero-based index of the outer interaction cycle containing success. Each generative-process step advances the MuJoCo plant by $\Delta t=0.002\,\mathrm{s}$; hence, the simulated time at success is $T_{\mathrm{sim}}=(T_{\mathrm{gp}}+1)\Delta t$. An outer cycle executes up to $\mathrm{AR}$ plant steps, with its final execution block truncated if success occurs. Neither timing metric measures wall-clock computation or HUA response latency. Timing summaries include successful episodes only. For the interaction-mediated condition, we additionally report the final posterior probability assigned to the true hypothesis, $q(z^\ast)$; posterior entropy, $H[q]$; MAP accuracy; and the frequency of $u_t^{\mathrm{preview}}=1$ HUA responses. These diagnostics are reported separately from task performance since exact latent-target identification and successful control under posterior-marginal preferences are distinct problems.


\section{Results}
\label{sec:results}

Table~\ref{tab:task-performance} reports task performance by condition and arm dimensionality. The direct-preference reference succeeded in $564/576$ episodes ($97.9\%$), compared with $423/576$ ($73.4\%$) under interaction-mediated control. The reference achieved higher success at every tested arm dimensionality. Among successful episodes, median time-to-first-success was longer under interaction mediation at every tested dimensionality.

\begin{table}[h]
\caption{Task performance by condition and arm dimensionality with matched LLCA settings. Success rates are proportions. Timing summaries include successful episodes only. $T_{\mathrm{gp}}$ is measured in generative-process steps ($\Delta t=0.002\,\mathrm{s}$ per step), and $T_{\mathrm{outer}}$ in outer interaction cycles. Each complete execution block spans $\mathrm{AR}$ plant steps, corresponding to $0.05\,\mathrm{s}$ for $N=3$ and $0.03\,\mathrm{s}$ for $N=4,5,6$.}
\label{tab:task-performance}
\centering
\scriptsize
\setlength{\tabcolsep}{2.5pt}

\begin{tabular}{cccccc|c|rcr|cc}
\toprule
Condition & $N_{links}$ & Plans & $H$ & AR & Episodes & Success & \multicolumn{3}{c|}{$T_{\mathrm{gp}}$} & \multicolumn{2}{c}{$T_{\mathrm{outer}}$} \\ 
 & & & & & & Rate & Mean & Median & IQR & Mean & Median  \\
\midrule
\multirow{4}{1.7cm}{\centering Direct-Preference LLCA} & 3 & 64 & 2 & 25 & 144 & 1.00 & 357.2 & 252 & 301.3 & 13.8 & 10 \\ 
& 4 & 64 & 2 & 15 & 144 & 1.00 & 249.3 & 177 & 197.3 & 16.1 & 11  \\ 
& 5 & 64 & 4 & 15 & 144 & 0.97 & 621.9 & 528 & 542.5 & 41.0 & 35  \\ 
& 6 & 32 & 4 & 15 & 144 & 0.95 & 883.8 & 683 & 777.0 & 58.4 & 45 \\ 
\midrule
\multirow{4}{1.7cm}{\centering Interaction-Mediated}& 3 & 64 & 2 & 25 & 144 & 0.67 & 1048.9 & 318 & 1154.0 & 41.5 & 12 \\ 
& 4 & 64 & 2 & 15 & 144 & 0.77 & 771.5 & 413 & 918.5 & 50.9 & 27  \\ 
& 5 & 64 & 4 & 15 & 144 & 0.72 & 804.4 & 568 & 824.5 & 53.2 & 37  \\ 
& 6 & 32 & 4 & 15 & 144 & 0.78 & 1047.5 & 774 & 1114.5 & 69.4 & 51 \\
\bottomrule
\end{tabular}
\end{table}

\begin{figure}[tbp]
    \centering
    \includegraphics[width=1.0\linewidth]{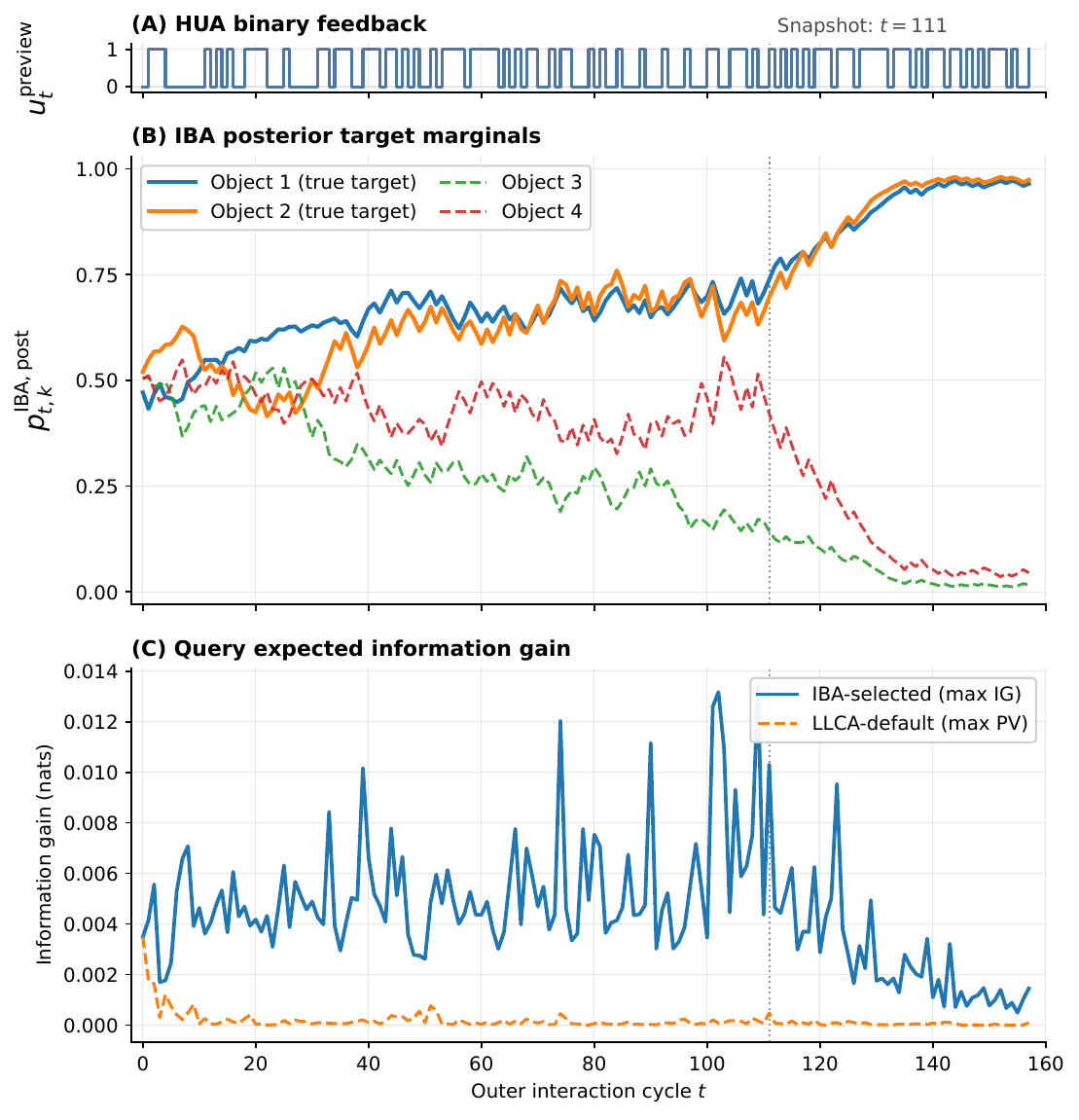}
    \caption{Temporal diagnostics show one illustrative, successful interaction-mediated episode. \textbf{(A)} Explicit HUA evaluations across outer cycles ($0$: satisfaction; $1$: dissatisfaction). \textbf{(B)} Post-feedback IBA target marginals; true-target labels identify HUA-private ground truth shown for offline diagnosis only. \textbf{(C)} Expected information gain of the IBA-selected (max-IG) and LLCA-default (max-PV) queries from the same pre-feedback rollout pool. The dotted line marks $t=111$, illustrated in Fig.~\ref{fig:example-fig-2}.}
    \label{fig:hua-iba-temporal-diagnostics}
\end{figure}

Table~\ref{tab:tripartite-diagnostics} reports final inference and interaction diagnostics. Exact target identification remained imperfect: MAP accuracy ranged from $0.167$ to $0.472$ across arm dimensionalities, while mean final posterior entropy remained non-zero. Figure~\ref{fig:hua-iba-temporal-diagnostics} illustrates the feedback and belief evolution within one illustrative, successful episode, not an aggregate trend. The mean dissatisfaction-response frequency ranged from $0.430$ to $0.468$. The LLCA acts on posterior-marginal preferences, not the MAP target subset; neither exact target identification nor posterior concentration is equivalent to achieving the coverage task within the episode horizon.

\begin{table}[t]
\caption{Interaction-mediated inference and feedback diagnostics by arm dimensionality; posterior statistics are evaluated at episode termination.}
\label{tab:tripartite-diagnostics}
\centering
\scriptsize
\setlength{\tabcolsep}{3pt}

\begin{tabular}{cc|c|cc|c|cc}
\toprule
$N_{links}$ & Episodes & HUA Dissatisfaction & \multicolumn{2}{c|}{$q(z^\ast)$} & MAP &  \multicolumn{2}{c}{$H[q]$} \\ 
 & & Rate Mean & Mean & Median & Accuracy& Mean & Median \\
\midrule
3 & 144 & 0.468 & 0.225 & 0.176 & 0.292 & 1.611 & 1.699 \\ 
4 & 144 & 0.449 & 0.337 & 0.241 & 0.472 & 1.428 & 1.638  \\ 
5 & 144 & 0.435 & 0.179 & 0.174 & 0.250 & 1.678 & 1.721  \\ 
6 & 144 & 0.430 & 0.182 & 0.169 & 0.167 & 1.695 & 1.730 \\ 
\bottomrule
\end{tabular}

\end{table}


\section{Discussion and Conclusion}
\label{sec:discussion}

Interaction-mediated control succeeded in $423/576$ episodes ($73.4\%$), compared with $564/576$ ($97.9\%$) under direct preference access. This difference plausibly reflects the additional challenge of inferring the task objective, although the experiment does not isolate the contributions of inference uncertainty, query selection, and control. The architecture achieved successful control across all tested arm dimensionalities despite imperfect target-subset identification. Since the LLCA acts on posterior-marginal object preferences rather than a MAP target hypothesis, exact preference identification and task completion constitute distinct experimental outcomes.

Our present implementation employs a simulated user whose feedback process \textit{matches} the response-model family assumed by the mediator. Consequently, the experiments do not establish robustness to human-model mismatch or actual human interaction behaviour. This matched-model assumption was deliberately adopted to evaluate an initial realisation of the proposed architecture under controlled conditions. The experimental task is restricted to a fixed object layout and target cardinality, while the controller employs calibrated random-shooting planning and the mediator uses a single expected-information-gain query-selection strategy. Furthermore, the simulation treats query selection and user feedback as instantaneous, without modelling human response latency or the real-time consequences of pausing plant execution during interaction. Future work should investigate user--mediator model mismatch, alternative interaction and query-selection mechanisms, and more computationally efficient planning methods for the low-level controller. Human-in-the-loop experiments and evaluations across other interactive systems will be necessary to establish the practical applicability and broader generality of the proposed architecture.

We have formulated and implemented a distributed Active Inference architecture that separates user-preference inference from autonomous plant control. In a simulated whole-arm coverage task, binary evaluative interaction enabled successful control without user-specified joint torques, albeit with lower success than direct preference access. These initial findings motivate further evaluation with human users and more varied interactive systems.

\subsubsection*{Acknowledgements.}
This work was funded by the Designing Interaction Freedom via Active Inference (DIFAI) ERC Advanced Grant (101097708), funded under the UK Horizon guarantee as EPSRC EP/Y029178/1. R.M-S. and J.H.W. acknowledge EPSRC EP/R018634/1; R.M-S. additionally acknowledges EP/T00097X/1 and EP/T021020/1. M.K. acknowledges the AI-Twin ERC Advanced Grant (101200584), funded by HORIZON.1.1--European Research Council (ERC).

\appendix
\clearpage

\section*{Appendix}

\section{Human/User Agent Implementation Details}
\label{app:hua}

\subsubsection{HUA Variables and Interface}
The simulated Human/User Agent (HUA) privately knows the true target indicator $z^\ast=(z_1^\ast,\ldots,z_K^\ast)\in\{0,1\}^K$, where $z_k^\ast=1$ iff object $k$ is a target. Its effective object-preference vector is therefore $p^{\mathrm{HUA}}=z^\ast$, which is not directly observable by either the IBA or LLCA. For LLCA candidate rollout $m$, define the terminal object-wise basis terms $B_{t,m,k}^{\mathrm{term}} := B_{t,m,H,k}^{\mathrm{step}}$ and collect all $M$ candidates into
\begin{equation}
    B_{t,\mathrm{all}}
    :=
    \left(
        B_{t,m,k}^{\mathrm{term}}
    \right)_{m=0,k=1}^{M-1,K}
    \in\mathbb{R}^{M\times K}.
\end{equation}

The HUA receives $B_{t,\mathrm{all}}$ in addition to the IBA-selected query index $m_t^{\mathrm{query}}$. Although the corresponding predicted rollout geometry may be visualised as the preview, the implemented response model is conditioned on the full basis pool and selected index rather than on the selected terminal-basis vector alone. The public response $u_t^{\mathrm{preview}} \in \{0,1\}$ denotes satisfaction ($u=0$) or dissatisfaction ($u=1$) with the selected preview \textit{relative to the current rollout-pool}. The absence of a ``dissatisfaction'' response ($u=1$) is interpreted as ``satisfaction'' ($u=0$), rather than a missing observation.

\subsubsection{HUA Feedback Process}
For target hypothesis $z$ and candidate $m$, define the target-conditioned terminal value
\begin{equation}
    V_t(z,m)
    =
    \sum_{k=1}^{K}
    z_k B_{t,m,k}^{\mathrm{term}}.
\end{equation}
Let $\mu_t(z)$ and $s_t(z)$ denote the population mean and standard deviation of $\{V_t(z,j)\}_{j=0}^{M-1}$ across the complete candidate pool, respectively. The pool-relative value of candidate $m$ is
\begin{equation}
    \zeta_t(z,m)
    =
    \frac{V_t(z,m)-\mu_t(z)}
         {s_t(z)+\epsilon},
\end{equation}
and its probability of eliciting a satisfaction response is
\begin{equation}
    r_t(z,m)
    :=
    \sigma\!\left(\alpha\,\zeta_t(z,m)\right),
\end{equation}
where $\epsilon>0$ provides numerical stability and $\alpha>0$ controls response steepness. Dependence on $B_{t,\mathrm{all}}$ is implicit through the rollout pool statistics. The public feedback likelihood is then
\begin{equation}
    P_B
    \left(
        u_t^{\mathrm{preview}}
        \mid
        z,B_{t,\mathrm{all}},m
    \right)
    =
    r_t(z,m)^{1-u_t^{\mathrm{preview}}}
    \left(1-r_t(z,m)\right)^{u_t^{\mathrm{preview}}}.
\end{equation}
The simulated HUA samples this likelihood at the privately known $z=z^\ast$ and selected query $m=m_t^{\mathrm{query}}$. Candidates that are comparatively favourable to $z^\ast$ within the current rollout pool are therefore more likely to elicit satisfaction. The HUA and IBA use the same pool-contextual response-model family, with the HUA evaluation at $z^\ast$ and the IBA under alternative latent target hypotheses (matched-model assumption). The HUA is a stochastic evaluative-feedback process and performs no VFE minimisation, EFE optimisation, or temporal planning.


\section{Interaction Broker Agent Implementation Details}
\label{app:iba}

\subsection{Latent Intent Model and Query Selection}
The Interaction Broker Agent (IBA) treats the HUA's private target subset as a latent variable $z \in \mathcal{Z}$, where for fixed target cardinality $K_{\mathrm{tar}}$, we have:
\begin{equation}
    \mathcal{Z}
    :=
    \left\{
        z\in\{0,1\}^K:
        \sum_{k=1}^{K}z_k=K_{\mathrm{tar}}
    \right\}.
\end{equation}
At outer cycle $t$, the IBA maintains a categorical pre-query belief $q_t^{\mathrm{pre}}(z)$ over $\mathcal{Z}$ and models prospective HUA feedback using the pool-contextual likelihood $P_B(u\mid z,B_{t,\mathrm{all}},m)$ defined in Appendix~\ref{app:hua}. In the reported experiments, $K=4$ and $K_{\mathrm{tar}}=2$, giving six hypotheses with uniform initial belief $q_0^{\mathrm{pre}}(z)=1/6$. For candidate query $m$, the predictive feedback distribution is
\begin{equation}
    P_t^{\mathrm{pred}}
    \left(
        u\mid B_{t,\mathrm{all}},m
    \right)
    =
    \sum_{z\in\mathcal{Z}}
    q_t^{\mathrm{pre}}(z)
    P_B
    \left(
        u\mid z,B_{t,\mathrm{all}},m
    \right).
\end{equation}
The IBA evaluates the mutual information between latent target hypothesis and prospective, binary HUA feedback directly as
\begin{align}
    \mathrm{IG}_t^{\mathrm{IBA}}(m)
    &=
    \sum_{z\in\mathcal{Z}}
    q_t^{\mathrm{pre}}(z)
    \sum_{u\in\{0,1\}}
    P_B
    \left(
        u\mid z,B_{t,\mathrm{all}},m
    \right)
    \nonumber\\
    &\qquad\times
    \log
    \frac{
        P_B
        \left(
            u\mid z,B_{t,\mathrm{all}},m
        \right)
    }{
        P_t^{\mathrm{pred}}
        \left(
            u\mid B_{t,\mathrm{all}},m
        \right)
    },
\end{align}
where
\begin{equation}
    m_t^{\mathrm{query}}
    =
    \arg\max_{m\in\{0,\ldots,M-1\}}
    \mathrm{IG}_t^{\mathrm{IBA}}(m).
\end{equation}
In consequence, query selection is purely epistemic in the present implementation; no separate pragmatic term enters the IBA objective.

\subsection{Posterior Inference and Preference Communication}
After observing the HUA's evaluative response: $u_t^{\mathrm{preview}}$, the IBA performs exact categorical Bayesian inference:

\begin{align}
    \ell_t(z)
    &=
    P_B
    \left(
        u_t^{\mathrm{preview}}
        \mid
        z,B_{t,\mathrm{all}},m_t^{\mathrm{query}}
    \right),
    \\
    Z_t
    &=
    \sum_{z'\in\mathcal{Z}}
    q_t^{\mathrm{pre}}(z')\ell_t(z'),
    \\
    q_t^{\mathrm{post}}(z)
    &=
    \frac{
        q_t^{\mathrm{pre}}(z)\ell_t(z)
    }{
        Z_t
    }.
\end{align}

Hence, the IBA updates its categorical belief via a direct application of Bayes' rule over the enumerated hypothesis space. The IBA does not perform VFE minimisation for this update. The raw categorical posterior $q_t^{\mathrm{post}}(z)$ is not sent to the LLCA. Instead, the IBA forms the object-wise marginal vector as defined in Eq.~\ref{eq:iba_target_beliefs}. Thus $p_{t,k}^{\mathrm{IBA},\mathrm{post}} = P_{q_t^{\mathrm{post}}}(z_k=1)$ is the posterior probability that object $k$ belongs to the HUA's target subset. These marginals, rather than the MAP hypothesis or full categorical posterior, parameterise the LLCA's object-wise preferences. After execution, the IBA's posterior is carried forward as $q_{t+1}^{\mathrm{pre}}(z)=q_t^{\mathrm{post}}(z)$.


\section{Low--Level Control Agent Implementation Details}
\label{app:llca}

\subsection{Generative Model: States, Observations, and Actions}
The Low--Level Control Agent (LLCA) is the exclusive controller of the articulated arm. It does not infer the HUA's latent target subset or model the HUA--IBA interaction. Instead, it receives an object-indexed preference vector from the IBA and determines the joint-torque actions required to realise those preferences. The true task environment is a planar $N$-link arm simulated by the MuJoCo/dm-control generative process. For the implementation, the physical angular configuration $\theta_t\in\mathbb{T}^N$ is represented by its MuJoCo joint-coordinate vector in $\mathbb{R}^N$, which we denote by the same symbol when discussing the implementation. We therefore write the generative process (plant) state in implemented coordinates as
$s_t^\top
=
\begin{bmatrix}
    \theta_t^\top &
    \dot{\theta}_t^\top
\end{bmatrix}
\in \mathbb{R}^{1 \times 2N}$
where $\theta_t \in \mathbb{R}^N$ and $\dot{\theta}_t \in \mathbb{R}^N$ are the joint-angle and joint-angular-velocity vectors, respectively. The LLCA acts directly on the plant through joint torques $a_t \equiv \tau_t \in \mathbb{R}^{N}$. The exact MuJoCo plant transition is treated as a generative-process black box and is not assumed to be available to the LLCA. The LLCA instead maintains an internal torque-driven dynamical model with latent state
$\hat{s}_t^\top
=
\begin{bmatrix}
    \hat{\theta}_t^\top &
    \hat{\dot{\theta}}_t^\top
\end{bmatrix}
\in
\mathbb{R}^{1 \times 2N}$.

The corresponding control variable in the LLCA's internal model is $\hat{\tau}_t\in\mathbb{R}^N$. Let $\hat{L}_k$ and $\hat{m}_k$ denote the agent's model of the length and mass of link $k$, respectively. Because $\hat{\theta}_1$ gives the orientation of the first link and each subsequent joint angle is relative to the preceding link, the world-frame orientation of link $k$ is the sum over all $j = 1, ..., k$ joint angles. The LLCA uses standard planar serial-link kinematics for links with uniform mass. Let $\hat{J}_{X_k}(\hat{\theta})\in\mathbb{R}^{2\times N}$ denote the Jacobian of the position of the centre-of-mass of link $k$, with respect to the joint coordinates. Let $\hat{J}_{\omega_k}\in\mathbb{R}^{1\times N}$ denote the corresponding angular-velocity Jacobian, and let $\hat{I}_k=\frac{1}{12}\hat{m}_k\hat{L}_k^2$ denote the moment of inertia of link $k$ about its centre of mass. The configuration-dependent inertia matrix used by the LLCA's model is
\begin{equation}
    \hat{M}(\hat{\theta})
    =
    \sum_{k=1}^{N}
    \left(
        \hat{m}_k
        \hat{J}_{X_k}(\hat{\theta})^{\top}
        \hat{J}_{X_k}(\hat{\theta})
        +
        \hat{I}_k
        \hat{J}_{\omega_k}^{\top}
        \hat{J}_{\omega_k}
    \right).
\end{equation}

With joint-damping matrix $D$ and integration step $\Delta t$, the internal model applies a semi-implicit Euler update with implicit damping,
\begin{align}
    \hat{\dot{\theta}}_{t+1}
    &=
    \left(
        \hat{M}(\hat{\theta}_t)
        +
        \Delta t D
    \right)^{-1}
    \left(
        \hat{M}(\hat{\theta}_t)\hat{\dot{\theta}}_t
        +
        \Delta t\,\hat{\tau}_t
    \right)\\
    \hat{\theta}_{t+1}
    &=
    \hat{\theta}_t
    +
    \Delta t\,\hat{\dot{\theta}}_{t+1}.
\end{align}

Writing the resulting deterministic transition as $f(\hat{s}_t,\hat{\tau}_t)$, the LLCA transition model is
\begin{equation}
    p
    \left(
        \hat{s}_{t+1}
        \mid
        \hat{s}_t,\hat{\tau}_t
    \right)
    =
    \mathcal{N}
    \left(
        f(\hat{s}_t,\hat{\tau}_t),
        Q
    \right),
\end{equation}
where $Q=10^{-3}I_{2N}$ represents process uncertainty and model mismatch. The LLCA has an angle-only observation modality. The active plant adapter passes the MuJoCo joint-position coordinates to the LLCA unchanged: $o_t = \theta_t \in \mathbb{R}^{N}$. No stochastic observation noise is added to this generative-process observation stream. Joint angular velocities, Cartesian arm geometry, and task-state variables are omitted from the LLCA observation vector. The LLCA's internal observation map is therefore $h(\hat{s}_t) = \hat{\theta}_t$, with observation model
\begin{equation}
    p
    \left(
        o_t
        \mid
        \hat{s}_t
    \right)
    =
    \mathcal{N}
    \left(
        h(\hat{s}_t),
        \Sigma_{\text{obs}}
    \right).
\end{equation}
Thus, $\Sigma_{\text{obs}} =  \sigma_{\theta}^{2}I_N$ represents assumed observation uncertainty within the LLCA's generative model rather than noise injected into the plant observation stream. In the reported experiments, $\sigma_{\theta}=10^{-4}$ and hence $\Sigma_{\text{obs}}=10^{-8}I_N$.

\subsection{LLCA State Inference}
Here, $t$ indexes individual generative-process filtering steps. The LLCA maintains Gaussian beliefs over its latent continuous state. At simulation initialisation, the Gaussian mean is bootstrapped from the full, generative-process reset state $[\,\theta_0;\dot{\theta}_0\,]$; the reported experiments use zero initial joint velocities. Thereafter, joint angular velocities are absent from the LLCA observation stream and remain latent variables inferred through dynamical prediction and angle-based perceptual updates. Given the corrected Gaussian state belief from the preceding filtering step and the torque $\hat{\tau}_{t-1}$ applied during the intervening transition, the LLCA propagates that belief through its internal dynamics using an Unscented Transform (UT),
\begin{equation}
    \left(
        \bar{\mu}_t,
        \bar{\Sigma}_t
    \right)
    =
    \operatorname{UT}
    \left(
        f(\cdot,\hat{\tau}_{t-1}),
        \mu_{t-1},
        \Sigma_{t-1}
    \right),
\end{equation}
after incorporating process uncertainty as $\bar{\Sigma}_t \leftarrow \bar{\Sigma}_t + Q$. The resulting Gaussian predictive prior is denoted
$q_t^{\mathrm{LLCA}}(\hat{s}_t) = \mathcal{N}(\bar{\mu}_t,\bar{\Sigma}_t)$. Upon receiving observation $o_t$, the LLCA obtains the corrected Gaussian variational posterior $q_t^{\mathrm{LLCA}}(\hat{s}_t\mid o_t) = \mathcal{N}(\mu_t,\Sigma_t)$ by minimising the perceptual variational free energy defined in Eq.~\ref{eq:vfe}.

Unlike the IBA's direct categorical Bayesian update, this LLCA perceptual update is implemented as an explicit variational optimisation. The expected observation-surprisal term is estimated using reparameterised Monte Carlo samples from the Gaussian posterior. The Gaussian mean and Cholesky-factor parameters are optimised using Adam. All reported experiments use $50$ posterior samples, $25$ Adam optimisation steps per perceptual update, and a learning rate of $0.05$. During an action-repeat block, each repeated generative-process step is accompanied by an LLCA predictive update, a plant transition, and a variational perceptual correction using the resulting joint-angle observation. Thus, perceptual correction occurs at every repeated plant step rather than only at the end of the action-repeat block. Replanning occurs after the complete block.

\subsection{LLCA Rollout Generation}
For each rollout pool $\Pi_t$ defined in the main text, candidate torque vectors are sampled element-wise from the configured uniform proposal distribution. In all reported experiments, candidate torque values are drawn from $[-1.5,1.5]$. This sampling process is independent of the object-wise preference vector $\mathbf{p}_t$. The latter affects the LLCA's preference-weighted rollout scores rather than the sampled torque sequences themselves. Each planned torque vector is propagated internally for $\texttt{action\_repeat}$ model steps. Thus, a candidate plan of horizon $H$ has effective model-prediction depth $H_{\mathrm{eff}} = H \cdot \texttt{action\_repeat}$.

For each candidate rollout, the LLCA recursively propagates its current Gaussian state belief through the internal transition model. If $q_{t,h}^{(m)}(\hat{s})$ denotes the predictive belief associated with candidate $m$ at planning step $h$, the corresponding Gaussian belief is propagated under each planned torque $\hat{\tau}_{t,h}^{(m)}$ using the same Unscented-Transform prediction described above. This propagation is repeated internally according to $\texttt{action\_repeat}$ before the rollout is scored at the next planning step. Each sampled plan therefore determines a predicted state-belief trajectory over the planning horizon. Rollout generation additionally produces an unweighted object-wise basis for each candidate plan. Let $B_{t,m,h,k}^{\mathrm{step}}$ denote the basis value for object $k$ at planning step $h$. Its terminal value is
\begin{equation}
    B_{t,m,k}^{\mathrm{term}}
    =
    B_{t,m,H,k}^{\mathrm{step}}.
\end{equation}

Collecting the terminal basis vectors for all $M$ candidate rollouts yields
$B_{t,\mathrm{all}}
\in
\mathbb{R}^{M\times K}.$
These unweighted basis values are computed from the deterministically reconstructed arm geometry, from the predicted joint-angle means and before application of the object-wise preference vector $\mathbf{p}_t$. Their precise coverage-based construction is specified below.

\subsection{LLCA Preference Model and Pragmatic Action Selection}
For either LLCA planning call, let $\mathbf{p}_t \in [0,1]^K$ denote the object-wise preference vector currently supplied to the LLCA. The LLCA converts these object-wise preference weights into an embodied whole-arm coverage preference. Recall from Section~\ref{sec:architecture} that $\mathcal{A}(\boldsymbol{\theta})$ denotes the geometry occupied by the whole arm under a joint-angle configuration. During LLCA rollout evaluation, the corresponding arm geometry is reconstructed deterministically from the joint-angle component of the predicted Gaussian mean. For an arbitrary joint-angle coordinate vector $\vartheta\in\mathbb{R}^N$ and task object $O_k$, with centre $c_k$ and radius $r_k$, define the minimum centre-to-arm distance as
\begin{equation}
    \delta_k(\vartheta)
    =
    \min_{i=1,\ldots,N}
    d
    \left(
        c_k,
        \mathcal{L}_i(\vartheta)
    \right).
\end{equation}

The penetration-adjusted distance is
$\bar{d}_k(\vartheta)
=
\max
\left(
    0,
    \delta_k(\vartheta)-r_k
\right)$,
such that $\bar{d}_k(\vartheta)=0$ iff $\mathcal{A}(\vartheta)$ intersects object $O_k$. We then define a smooth coverage feature as
\begin{equation}
    \phi_k(\vartheta)
    =
    \exp
    \left(
        -\frac{
            \bar{d}_k(\vartheta)^2
        }{
            2\sigma_b^2
        }
    \right).
\end{equation}
Collecting these features yields $\phi(\vartheta)\in[0,1]^K$, which is the coverage-feature vector entering the preference density in Eq.~\ref{eq:llca_preference}. For candidate rollout $\pi$, let $\mu_h^\pi\in\mathbb{R}^{2N}$ denote the mean of the predicted Gaussian state belief at planning step $h$, and let $\mu_{\theta,h}^\pi\in\mathbb{R}^{N}$ denote its joint-angle component. Rather than evaluating the expectation over the predicted state belief in Eq.~\ref{eq:pragmatic_value}, the reported implementation uses the following mean-state approximation,
\begin{equation}
    \mathbb{E}_{
        q^{\mathrm{LLCA}}(\hat{s}_h\mid\pi)
    }
    \left[
        \log
        P_C
        \left(
            \phi(\hat{\theta}_h)
            \mid
            \mathbf{p}_t
        \right)
    \right]
    \approx
    \log
    P_C
    \left(
        \phi(\mu_{\theta,h}^{\pi})
        \mid
        \mathbf{p}_t
    \right).
\end{equation}

This approximation evaluates the coverage preference at the predicted joint-angle mean and therefore neglects the contribution of predicted state covariance to the pragmatic value. For notational simplicity, we continue to denote the resulting implemented approximation by $\mathrm{PV}_h(\pi)$. The LLCA evaluates whole-arm coverage deterministically at this predicted joint-angle mean. The unweighted object-wise basis term for candidate $m$ is therefore
\begin{equation}
    B_{t,m,h,k}^{\mathrm{step}}
    =
    -
    \frac{1}{2\sigma_{\phi}^{2}}
    \left(
        1-
        \phi_k
        \left(
            \mu_{\theta,h}^{\pi_t^{(m)}}
        \right)
    \right)^2.
\end{equation}

The pragmatic value at planning step $h$ is obtained by weighting these object-wise basis terms by the current LLCA preference message,
\begin{equation}
    \mathrm{PV}_h
    \left(
        \pi_t^{(m)}
    \right)
    =
    \sum_{k=1}^{K}
    p_{t,k}
    B_{t,m,h,k}^{\mathrm{step}}.
\end{equation}

The LLCA implementation supports the pragmatic value/information-gain decomposition of the negative expected free energy (NEFE) planning objective,
\begin{equation}
    -G(\pi)
    \approx
    \frac{1}{H}
    \sum_{h=1}^{H}
    \left[
        \lambda_{\mathrm{pv}}\mathrm{PV}_h(\pi)
        +
        \mathrm{IG}_h(\pi)
    \right],
    \label{eq:efe}
\end{equation} 
where $\lambda_{\mathrm{pv}}\geq 0$ weights pragmatic value and $\mathrm{IG}_h(\pi)$ denotes the LLCA's prospective information-gain contribution, distinct from the IBA's information gain over user preferences. In all reported experiments, LLCA information gain is disabled and $\lambda_{\mathrm{pv}}=1$. Consequently, the implemented rollout-selection objective reduces to maximising mean whole-horizon pragmatic value, as specified in the main text. Only the first distinct torque vector of the newly selected execution rollout is used for plant control. The candidate planning range may exceed the physical actuator range; the plant therefore clips the selected torque component-wise to its actuator bounds. In the reported environment these bounds are $[-1,1]$, so the applied torque is
$\tau_t
=
\operatorname{clip}
\left(
    \hat{\tau}_{t,1}^{\star},
    -1,
    1
\right)$.

This torque vector is held fixed and repeatedly applied for the configured $\texttt{action\_repeat}$ generative-process steps. During each repeated step, the LLCA performs the filtering sequence described above: predictive propagation, plant transition, and variational perceptual correction. The remaining distinct torque vectors in the sampled rollout are discarded, and the LLCA replans after completion of the action-repeat block. 

\begin{figure}[h]
    \centering
    \includegraphics[width=\linewidth]{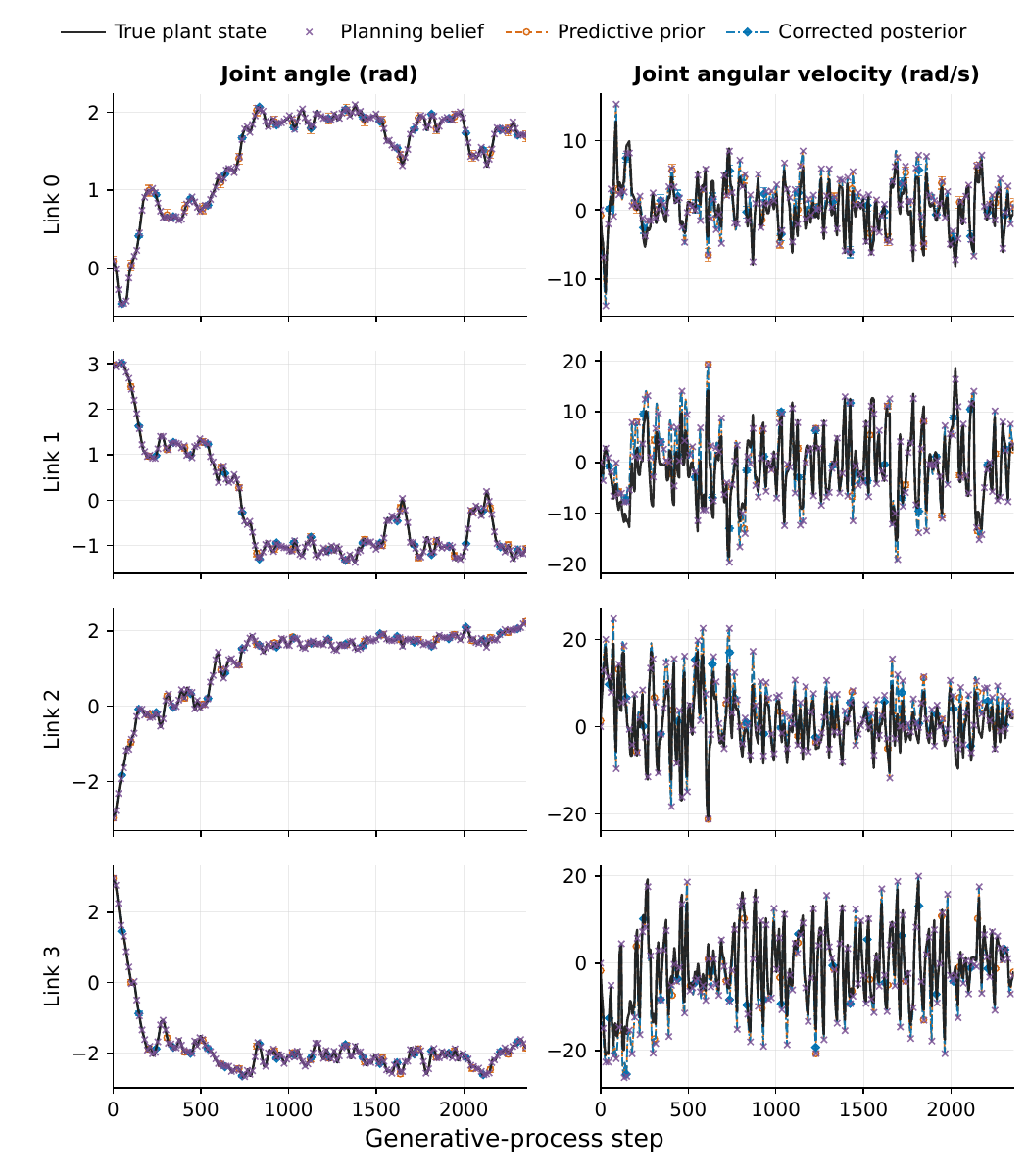}
    \caption{LLCA state-estimation trajectories for an illustrative four-link episode. Rows correspond to links 0--3; the left and right columns show joint angles and angular velocities, respectively. True plant states are compared with planning-time beliefs, predictive priors, and corrected posteriors. Error bars indicate $\pm 2$ standard deviations for the prior and posterior at selected filtering steps. Following initialisation, the LLCA observes joint angles only; angular velocities remain latent and are inferred through dynamical prediction and perceptual correction.}
    \label{fig:appendix-llca-posterior-marginals}
\end{figure}

\section{Complete Interaction Algorithm}
\label{app:interalg}

\begin{algorithm}[h]
\caption{One complete outer interaction cycle}
\label{alg:outer-cycle}
\begin{algorithmic}[1]
\Procedure{OuterInteractionCycle}{$q_t^{\mathrm{pre}}$}

    \State $\mathbf{p}_t^{\mathrm{IBA},\mathrm{pre}}
        \gets \mathbb{E}_{q_t^{\mathrm{pre}}}[z]$
        \Comment{pre-feedback preferences}

    \State $(\Pi_t^{\mathrm{query}},B_{t,\mathrm{all}})
        \gets \Call{LLCAPlan}{
            \mathbf{p}_t^{\mathrm{IBA},\mathrm{pre}}}$

    \State $m_t^{\mathrm{query}}
        \gets \arg\max_m \mathrm{IG}_t^{\mathrm{IBA}}(m)$

    \State $u_t^{\mathrm{preview}}
        \gets \Call{HUAFeedback}{
            B_{t,\mathrm{all}},m_t^{\mathrm{query}}}$
        \Comment{query only; not executed}

    \State $q_t^{\mathrm{post}}
        \gets \Call{IBAUpdate}{
            q_t^{\mathrm{pre}},
            u_t^{\mathrm{preview}},
            B_{t,\mathrm{all}},
            m_t^{\mathrm{query}}}$

    \State $\mathbf{p}_t^{\mathrm{IBA},\mathrm{post}}
        \gets \mathbb{E}_{q_t^{\mathrm{post}}}[z]$

    \State $\Pi_t^{\mathrm{exec}}
        \gets \Call{LLCAPlan}{
            \mathbf{p}_t^{\mathrm{IBA},\mathrm{post}}}$

    \State $\pi_t^\star
        \gets \Call{SelectByPV}{\Pi_t^{\mathrm{exec}}}$

    \State $\tau_t \gets
        \operatorname{clip}(\hat{\tau}_{t,1}^{\star},-1,1)$
    \For{$j=1,\ldots,\mathrm{AR}$}
        \State \Call{LLCAPredict}{$\hat{\tau}_{t,1}^{\star}$}
        \State \Call{PlantStep}{$\tau_t$}
        \State $o \gets \Call{ObserveJointAngles}{}$
        \State \Call{LLCACorrect}{$o$}
        \If{\Call{TaskSucceeded}{}}
            \State \Return \Comment{episode terminates}
        \EndIf
    \EndFor

    \State $q_{t+1}^{\mathrm{pre}}
        \gets q_t^{\mathrm{post}}$
        \Comment{carry belief forward}

\EndProcedure
\end{algorithmic}
\end{algorithm}

\FloatBarrier
\bibliographystyle{splncs04}
\bibliography{references}

\end{document}